\documentclass{sigchi-ext}
\usepackage[T1]{fontenc}
\usepackage{textcomp}
\usepackage[scaled=.92]{helvet} 
\usepackage{graphicx} 
\usepackage{balance}  
\usepackage{booktabs} 
\usepackage{ccicons}  
\usepackage{ragged2e} 

\copyrightinfo{Permission to make digital or hard copies of part or all of this work
   for personal or classroom use is granted without fee provided that
   copies are not made or distributed for profit or commercial advantage
   and that copies bear this notice and the full citation on the first
   page. Copyrights for third-party components of this work must be
   honored. For all other uses, contact the
   owner\hspace*{.5pt}/author(s).\\
   CSCW'18 Workshop on Industrial Internet of Things, Jersey City, New Jersey, USA
   }

\def\plaintitle{Hybrid Approach to Automation, RPA and Machine Learning: a Method for the Human-centered Design of Software Robots} 
\def\plainkeywords{robotic process automation; RPA, software robots; participatory design; co-programming; Industry 4.0; Living Lab; interactive and collaborative machine learning}

\title{Hybrid Approach to Automation, RPA and Machine Learning: a Method for the Human-centered Design of Software Robots}

\numberofauthors{9}
\author{%
  \alignauthor{%
    \textbf{Wies\l{}aw Kope\'{c}}\\
    \textbf{Kinga Skorupska}\\
    \textbf{Piotr Gago}\\
    \textbf{Krzysztof Marasek}\\
    \affaddr{Polish-Japanese Academy\\of Information Technology} \\
    \affaddr{Warsaw, Poland} \\
    \email{\{kopec,kinga.skorupska,\\pgago,kmarasek\}@pja.edu.pl}\\
    } 
    \alignauthor{%
    \textbf{Cezary Biele}\\
    \textbf{Katarzyna Abramczuk}\\
    \affaddr{National Information Processing Institute}\\
    \affaddr{Warsaw, Poland}\\
    \email{\{cezary.biele,\\katarzyna.abramczuk@opi.org.pl\}}
    } 
    \vfill
    \alignauthor{%
    \textbf{Marcin Skibi{\'n}ski}\\
    \affaddr{DPC Polska}\\
    \affaddr{Warsaw, Poland}\\
    \email{m.skibinski@dpcpolska.pl}
    }
    \alignauthor{%
    \textbf{Anna Jaskulska}\\
    \affaddr{Kobo Association}\\
    \affaddr{Warsaw, Poland}\\
    \email{a.jaskulska@kobo.org.pl} 
    }
    \vfill
    \alignauthor{%
    \textbf{Dominika Tkaczyk}\\
    \affaddr{University of Warsaw, ICM}\\
    \affaddr{Warsaw, Poland}\\
    \email{d.tkaczyk@uw.edu.pl} 
    }
}

\definecolor{linkColor}{RGB}{6,125,233}
\hypersetup{%
  pdftitle={\plaintitle},
  pdfkeywords={\plainkeywords},
  bookmarksnumbered,
  pdfstartview={FitH},
  colorlinks,
  citecolor=black,
  filecolor=black,
  linkcolor=black,
  urlcolor=linkColor,
  breaklinks=true,
}

\begin{document}

\CopyrightYear{2018}
\setcopyright{rightsretained}
\conferenceinfo{CSCW'18 Workshop on Industrial Internet of Things}{Jersey City, New Jersey, USA}


\maketitle

\RaggedRight{} 


\begin{abstract}

One of the more prominent trends within Industry 4.0 is the drive to employ Robotic Process Automation (RPA), especially as one of the elements of the Lean approach. The full implementation of RPA is riddled with challenges relating both to the reality of everyday business operations, from SMEs to SSCs and beyond, and the social effects of the changing job market. To successfully address these points there is a need to develop a solution that would adjust to the existing business operations and at the same time lower the negative social impact of the automation process. 

To achieve these goals we propose a hybrid, human-centered approach to the development of software robots.
This design and implementation method combines the Living Lab approach with empowerment through participatory design to kick-start the co-development and co-maintenance of hybrid software robots which, supported by variety of AI methods and tools, including interactive and collaborative ML in the cloud, transform menial job posts into higher-skilled positions, allowing former employees to stay on as robot co-designers and maintainers, i.e. as co-programmers who supervise the machine learning processes with the use of tailored high-level RPA Domain Specific Languages (DSLs) to adjust the functioning of the robots and maintain operational flexibility.

\end{abstract}





\marginpar{%
\vspace{40pt}
  \fbox{%
    \begin{minipage}{.975\marginparwidth}
      \vspace{0.25pc}
      \textbf{Key Challenges to \\Software Automation}\\
      \vspace{0.5pc} 
      \textbf{Technical:} \\
        {\small        
            \begin{itemize}[leftmargin=0.1in,noitemsep,topsep=0pt,parsep=0pt,partopsep=0pt]
                \item Costly and tedious maintenance
                \item Multiple input data formats
                \item Keeping paper documentation
                \item Low quality of existing data
            \end{itemize}
        }

      \vspace{0.5pc} \textbf{Organizational:} \\
        {\small
            \begin{itemize}[leftmargin=0.1in,noitemsep,topsep=0pt,parsep=0pt,partopsep=0pt]
                \item Lack of clear processes
                \item Multiple fragmentary solutions
                \item Unique legacy software
            \end{itemize}
        }
      
      \vspace{0.5pc} \textbf{Socioeconomic:} \\ 
        {\small
            \begin{itemize}[leftmargin=0.1in,noitemsep,topsep=0pt,parsep=0pt,partopsep=0pt]
                \item Loss of job posts
                \item Lack of awareness of RPA
            \end{itemize}
        }
      \vspace{0.5pc}
      
    \end{minipage}}
    \label{sec:sidebar1} 
}

\section{Introduction}

As Economy 4.0 is firmly upon us the digitization and mechanization of business processes is increasing in all industries. Thus, to stay competitive it is no longer enough to optimize and simplify processes but there is a need to support them with automation, especially if they are repetitive, predictable and mostly digital.

Within Industry 4.0 there exists a prominent trend to employ Robotic Process Automation (RPA) especially 
as one of the elements of the Lean approach. To follow it companies often need to adjust the industry processes to the limitations of bulky, automated, often rule-based, IT systems and the format and structure of the input and output data required. Apart from the need to transform the business operations, there is also the negative impact this shift may have on the job market within a given industry, as the target job posts are fully reduced due to automation.

To achieve these goals we propose a hybrid, human-centered approach to the development of software robots.
This design and implementation method combines the Living Lab approach\cite{ogonowski2013designing, niitamo2006state} with empowerment through participatory design to kick-start the co-development of hybrid software robots which,
supported by variety of AI methods and tools, including interactive and collaborative ML\cite{robert2016reasoning}, transform menial job posts into higher-skilled positions, allowing some of the former employees to stay on as robot
co-programmers and co-maintainers who supervise machine learning and use pseudo-code to adjust the functioning of the robots to maintain operational flexibility. We believe that this approach will provide a relatively low-cost and user-friendly RPA implementation and sustainable maintenance solution.

\section{Challenges to the Implementation of Automated Processes}

The use of Robotic Process Automation is on the rise throughout the Western economies
 as the automation it offers can solve one of the largest business challenges of today: the need to process increasingly larger amount of data.

However, this method of automation itself since its inception has faced multiple challenges which still remain relevant. They are concentrated in three key areas depicted in the sidebar note
related to technical, organizational and socioeconomic aspects. 

\subsection{Technical challenge}

Many rule-based robots are difficult to scale because the rules are written by hand. Moreover, they are difficult to maintain to remain flexible, given the varied format and structure of the data to be processed, which often includes e-mails, web forms, faxes, scans of paper documents, phone calls or even financial or sensor data. Moreover, some output documents still need to be produced in paper and sent out to clients. Data already present in the current systems is often of poor quality in general, or for automation as it lacks tags and division by categories; on top of this it may be outdated and based on old regulations, checks and processes which makes it difficult to use machine learning to properly train neural networks.

\marginpar{%
\vspace{-45pt}
  \fbox{%
    \begin{minipage}{.975\marginparwidth}
      \vspace{0.25pc}
      \textbf{Benefits of the Hybrid \\Approach}\\
      \vspace{0.5pc} 
      {\small 
      \textbf{1. Flexibility and Participatory Maintenance :}  AI and ML-Powered design allows for automation is derived from the patterns taught by employees in a Living Lab environment, who in turn learn how to facilitate this process.
        }

      \vspace{0.5pc} 
      {\small 
      \textbf{2. Adjustment to Existing Business Processes:} This ensures lower entry barrier than classic BPA with process re-engineering. The businesses can retain their current practices, supplementing human work with RPAs where possible, which means that the entry barrier is lower, as this solution is cheaper than BPA with process re-engineering and seamless, allowing for continuity of operation.
      }
      
      \vspace{0.5pc} 
      {\small 
      \textbf{3. Empowerment:} Some employees who used to perform low-skilled jobs instead of being displaced, become co-creators and co-maintainers of the software robots.
      }
      \vspace{0.5pc}
      
    \end{minipage}}
    \label{sec:sidebar2} 
}

\subsection{Organizational challenge}

In multiple companies the existing business processes rely on complex chains of manufacturing involving multiple approval steps, contractors, clients and convoluted internal procedures. Such companies often use legacy software, which was developed with the rise of new business needs. These custom solutions are often rule-based and maintained by different external contractors. Such software can be fragmentary in nature as different processes are segmented into different software solutions, which often are not compatible with either the current standards, or even one another and can only be maintained through constant patching.
When automation is delivered with RPA solutions, they are often neither intuitive nor user-friendly, and prone to errors. At the same time, multiple organizations lack sufficient knowledge about their own business processes, especially on handling exceptions and allowing for shortcuts and bypasses to effectively build such robots on their own.

\subsection{Socioeconomic challenge}

The implementation of automation with software robots is connected to ethical dilemmas as it is often followed by restructuring and massive loss of job posts\cite{frey2017future}. This general awareness is one of the barriers to automation, as the managing staff and employees are aware of the associated risks and benefits, but have limited knowledge of how to mitigate them.

\section{Discussion of the Proposed Solutions}

The three problem areas discussed above can be directly addressed by a shift in thinking about Robotic Process Automation from a purely algorithmic IT perspective to the HCI one. Ultimately, RPA can become a human-centered endeavor as software robots relieve employees of tedious repetitive tasks, allowing them to increase their competences and build value in other areas. Thus, below we discuss the proposed solutions.

\subsection{ Distributed and Crowdsourced Machine Learning Approach}

This postulate coresponds with the novel interactive and collaborative apprach to machine learning\cite{robert2016reasoning}. In particular we think that the use of neural networks to expand the functionality of software robots can ensure that various input and output data is efficiently handled, including audio and images. This process ought to be supported by empowered employees, who can verify the quality of the input and output data retrived and analysed by various ML tools and techniques \cite{tkaczyk2015cermine, tkaczyk2015structured}, including correct OCR-tagging, handling exceptions and rare cases. This process should be supported by state-of-the-art technology based on our advanced research on eyetracking methods and techniques\cite{biele2018eye, biele2018surface}. Finally, the employees from co-maintainers of the solution can become co-designers and
co-programmers, as they learn to modify the pseudo-code responsible for the functioning of the software robot they oversee.

\subsection{Supplementation of Existing Solutions}

The use of software robots which directly emulate the jobs of specific human employees allows the companies to continue to use their time-tested business processes and legacy software. Moreover, the central platform for RPA and the high level language tailored to the crafted RPA Domain Specific Language are easier to maintain, and will remain up to date, including the cloud-based Machine Learning and Neural Network components employed in the abovementined interactive and collaborative mode\cite{robert2016reasoning}. The 1:1 mapping of the human tasks and the involvement of current employees as co-programmers ensures the continuity of business operations as exceptions can be handled on the fly by internal staff.

\subsection{Employee Empowerment through Participation}

The implementation of software robots need not be followed by massive layoffs. Based on our previous advances in participatory design and co-design\cite{kopec2017living, kopec2017spiral}, Living Lab activities\cite{kopec2017living} and higher level crowdsourcing solutions coupled with cloud-based collaborative solutions for quality assurance, also in ML\cite{skorupska2018smarttv}, we postulate that, according to the principles of Participatory Design, employees who work in the target capacity of the robots can be the best co-designers and co-maintainers. Through their empowerment via participation in Living Labs and training they are motivated to increase their competences and learn how to efficiently perform their job of supervisors of machine learning and hybrid RPA high-level DLS programmers. 
Through this their jobs are transformed from menial to skilled, and their time is freed to work on more challenging aspects of the business.

\section{A Holistic Solution: The Hybrid Approach}

Thus, we propose the creation of a software platform and hardware setup that would allow employees to participate in the development and maintenance of software robots supported by various AI-powered solutions, including neural networks and interactive and collaborative machine learning. This solution ensures business continuity as the learning process and software robot operations are overseen by employees possessing the internal know-how of the company. This platform would allow businesses to retain the most motivated of their current employees while at the same time empowering them to learn new skills of co-maintainers and co-programmers. The implementation process of this Hybrid Approach can consist of the following stages:

\begin{enumerate}
\item \textbf{Analysis of RPA penetration and potential within the company}
\item \textbf{Workshops with employees to identify opportunity areas}
\item \textbf{Living Lab approach to process analysis with data-collection workstations}
\item \textbf{Participatory design of specific software robots}
\item \textbf{Supervised Training of AI-based solutions}
\item \textbf{Employee empowerment training sessions}
\item \textbf{Co-programming and co-maintenance of software robots in RPA DSLs}
\end{enumerate}

\section{Conclusions}

As the trends in Industry 4.0 are leaning towards automation there the need to develop positive human-centered methods. Thus, in this abstract we proposed a design approach that addresses key challenges of RPA. The method relies on participatory design of software robots, facilitated by a Living Lab environment, interactive and collaborative AI solutions, including machine learning and neural networks whereby menial tasks are turned into high-skilled jobs, increasing employee satisfaction and lowering turnover usable in multiple contexts, such as for example automatic tests, code deployment, customer service and machine and device control.

\bibliography{bibliography}


\begin{thebibliography}{00}


\ifx \showCODEN    \undefined \def \showCODEN     #1{\unskip}     \fi
\ifx \showDOI      \undefined \def \showDOI       #1{{\tt DOI:}\penalty0{#1}\ }
  \fi
\ifx \showISBNx    \undefined \def \showISBNx     #1{\unskip}     \fi
\ifx \showISBNxiii \undefined \def \showISBNxiii  #1{\unskip}     \fi
\ifx \showISSN     \undefined \def \showISSN      #1{\unskip}     \fi
\ifx \showLCCN     \undefined \def \showLCCN      #1{\unskip}     \fi
\ifx \shownote     \undefined \def \shownote      #1{#1}          \fi
\ifx \showarticletitle \undefined \def \showarticletitle #1{#1}   \fi
\ifx \showURL      \undefined \def \showURL       #1{#1}          \fi

\bibitem{biele2018surface}
{Cezary Biele} {and} {Pawel Kobylinski}. 2018.
\newblock \showarticletitle{Surface Recalibration as a New Method Improving
  Gaze-Based Human-Computer Interaction}. In {\em International Conference on
  Intelligent Human Systems Integration}. Springer, 197--202.
\newblock


\bibitem{biele2018eye}
{Andrew~T. Duchowski}, {Krzysztof Krejtz}, {Izabela Krejtz}, {Cezary Biele},
  {Anna Niedzielska}, {Peter Kiefer}, {Martin Raubal}, {and} {Ioannis
  Giannopoulos}. 2018.
\newblock \showarticletitle{The Index of Pupillary Activity: Measuring
  Cognitive Load Vis-\`{a}-vis Task Difficulty with Pupil Oscillation}. In {\em
  Proc. of the 2018 CHI Conf. on Human Factors in Computing Systems} {\em (CHI
  '18)}. ACM, NY, USA, Article 282.
\newblock
\showISBNx{978-1-4503-5620-6}


\bibitem{frey2017future}
{Carl~Benedikt Frey} {and} {Michael~A Osborne}. 2017.
\newblock \showarticletitle{The future of employment: how susceptible are jobs
  to computerisation?}
\newblock {\em Technological forecasting and social change\/}  {114} (2017),
  254--280.
\newblock


\bibitem{kopec2017spiral}
{Wies{\l}aw Kope{\'c}}, {Rados{\l}aw Nielek}, {and} {Adam Wierzbicki}. 2018.
\newblock \showarticletitle{Guidelines Towards Better Participation of Older
  Adults in Software Development Processes using a new SPIRAL Method and
  Participatory Approach}. In {\em Proceedings of the CHASE'18: International
  Workshop on Cooperative and Human Aspects of Software} {\em (ICSE '18)}. ACM,
  New York, NY, USA.
\newblock
\showDOI{%
\url{http://dx.doi.org/10.1145/3195836.3195840}}


\bibitem{kopec2017living}
{Wies{\l}aw Kope\'{c}}, {Kinga Skorupska}, {Anna Jaskulska}, {Katarzyna
  Abramczuk}, {Radoslaw Nielek}, {and} {Adam Wierzbicki}. 2017.
\newblock \showarticletitle{LivingLab PJAIT: Towards Better Urban Participation
  of Seniors}. In {\em Proceedings of the International Conference on Web
  Intelligence} {\em (WI '17)}. ACM, New York, NY, USA, 1085--1092.
\newblock
\showISBNx{978-1-4503-4951-2}
\showDOI{%
\url{http://dx.doi.org/10.1145/3106426.3109040}}


\bibitem{niitamo2006state}
{Veli-Pekka Niitamo}, {Seija Kulkki}, {Mats Eriksson}, {and} {Karl~A
  Hribernik}. 2006.
\newblock \showarticletitle{State-of-the-art and good practice in the field of
  living labs}. In {\em Technology Management Conference (ICE), 2006 IEEE
  International}. IEEE, 1--8.
\newblock


\bibitem{ogonowski2013designing}
{Corinna Ogonowski}, {Benedikt Ley}, {Jan Hess}, {Lin Wan}, {and} {Volker
  Wulf}. 2013.
\newblock \showarticletitle{Designing for the living room: long-term user
  involvement in a living lab}. In {\em Proc. of the SIGCHI Conference on Human
  Factors in Computing Systems}. ACM, 1539--1548.
\newblock


\bibitem{robert2016reasoning}
{Sebastian Robert}, {Sebastian B{\"u}ttner}, {Carsten R{\"o}cker}, {and}
  {Andreas Holzinger}. 2016.
\newblock \showarticletitle{Reasoning under uncertainty: Towards collaborative
  interactive machine learning}.
\newblock In {\em Machine learning for health informatics}. Springer, 357--376.
\newblock


\bibitem{skorupska2018smarttv}
{Kinga Skorupska}, {Manuel N\'{u}\~{n}ez}, {Wies{\l}aw Kope{\'c}}, {and}
  {Radoslaw Nielek}. 2018.
\newblock \showarticletitle{Older Adults and Crowdsourcing: Android TV App for
  Evaluating TEDx Subtitle Quality}.
\newblock {\em Proc. ACM Hum.-Comput. Interact.\/} {2}, CSCW, Article 159 (Nov.
  2018), 23 pages.
\newblock
\showISSN{2573-0142}
\showDOI{%
\url{http://dx.doi.org/10.1145/3274428}}


\bibitem{tkaczyk2015cermine}
{Dominika Tkaczyk}, {Pawe{\l} Szostek}, {Mateusz Fedoryszak}, {Piotr~Jan
  Dendek}, {and} {{\L}ukasz Bolikowski}. 2015a.
\newblock \showarticletitle{CERMINE: automatic extraction of structured
  metadata from scientific literature}.
\newblock {\em International Journal on Document Analysis and Recognition
  (IJDAR)\/} {18}, 4 (01 Dec 2015), 317--335.
\newblock
\showISSN{1433-2825}
\showDOI{%
\url{http://dx.doi.org/10.1007/s10032-015-0249-8}}


\bibitem{tkaczyk2015structured}
{Dominika Tkaczyk}, {Bartosz Tarnawski}, {and} {Lukasz Bolikowski}. 2015b.
\newblock \showarticletitle{Structured affiliations extraction from scientific
  literature}.
\newblock {\em D-Lib Magazine\/} {21}, 11/12 (2015).
\newblock
\showDOI{%
\url{http://dx.doi.org/10.1045/november2015-tkaczyk}}


\end{thebibliography}
\bibliographystyle{SIGCHI-Reference-Format}

\end{document}